\documentstyle[12pt]{article}

\newcommand{\beq}{\begin{equation}}
\newcommand{\eeq}{\end{equation}}
\newcommand{\barr}{\begin{eqnarray}}
\newcommand{\earr}{\end{eqnarray}}
\newcommand{\ba}{\begin{array}}
\newcommand{\ea}{\end{array}}

\begin{document}
\begin{center}
{\bf \large SUM RULES AND POSITIVITY CONSTRAINTS\\
ON NUCLEON SPIN STRUCTURE}

\vspace*{2cm}

{\it  O.TERYAEV$^{a,b}$, B.PIRE$^a$, J.SOFFER$^c$}

\vspace*{2cm}

$^a$CPhT\footnote {Unit\'e Mixte de Recherche
C7644 du Centre National de la Recherche
Scientifique}, Ecole Polytechnique, F-91128
Palaiseau, France\\
\vskip 0.5cm
$^b$Bogoliubov Laboratory of Theoretical Physics,\\
Joint Institute for Nuclear Research\\
141980 Dubna, Moscow region, Russia\footnote{Permanent address} \\
\vskip 0.5cm
$^c$CPT-CNRS\footnote {Unit\'e Propre de Recherche 7061 du
Centre National de la Recherche Scientifique}, F-13288 Marseille Cedex
9, France
\end{center}

{\bf Abstract}.
 The spin structure of nucleon at twist 2 and 3
levels is analyzed. The contribution of quark and gluon spins to 
nucleon spin are Lorentz invariant, while it is not sure for
orbital angular momentum. The conserved fractional moments
of transversity distribution are considered. The scenario
of decoupled total angular momentum, determined predominantly by the
unpolarized scattering, is  discussed.

\vspace*{2cm}

\section{Introduction}
The spin structure of nucleons is still one of the major puzzles of
hadronic physics. Ten years after the first experimental data on
polarized structure functions opened the {\it spin crisis}, much
progress has been achieved in the QCD understanding of sum rules, $Q^2$
evolution of distribution functions and the orbital angular momentum
contribution to the total spin\cite{Review}. Moreover,
transversity\cite{trans}  and off-forward\cite{Ji,OFPD} distribution
functions now appear as complementary sources of information. In this
paper, we focus on spin sum rules and on positivity constraints which
have been recently derived. We also address questions related to the
contribution  of orbital angular momentum to the total spin.

\section{Spin sum rule}

The fact that the total nucleon spin is just $1/2$ is
usually expressed as a sum rule

\begin{equation}\label{SR}
J_q+J_G=
{1 \over 2} \sum_{q, \bar q} S_q + S_G+L_q+L_G= {1 \over 2},
\eeq

\noindent
for the first moments $S$ of quark and gluon spin-dependent
distributions.

\begin{equation}\label{S}
S_q=\int_0^1 \Delta q(x) dx; ~~~~~~ S_G=\int_0^1 \Delta G(x) dx.
\eeq

The orbital angular momentum $L$ is a necessary ingredient
to make the sum rule (\ref{SR}) $Q^2$ independent \cite{Ratcl}.
The different roles of $S$ and $L$ entering the
sum rule (\ref{SR}) are manifested by the fact that the spin parts
(\ref{S})   are naturally  measured in inclusive processes like deep
inelastic  scattering or high-$p_T$ direct photon production, while
indirect access to $L$ requires exclusive reactions like deeply virtual
Compton scattering\cite{Ji}, with essential experimental and theoretical
difficulties accompanying its extraction from the data\cite{DGPR}.

The possible resolution of this paradox might be the effective
decoupling \cite{OT98} of the orbital and total angular momenta,
resulting in the equal sharing of the momentum and total angular
momentum of quarks and gluons. This would naturally explain the
observed smallness of the isoscalar anomalous magnetic moment
($1.79=\mu^A_p \sim - \mu_n^A =1.91$), which is also manifested 
in the model calculations \cite{Man}. 
The reason is that the anomalous magnetic moment 
related to the non-forward distribution $E$ \cite{Ji}, leading the
a different sharing of momentum and total angular
momentum. The recently suggested 
representation of the quark total angular momentum 
\cite{Ji3},
\begin{equation}\label{Lq}
J_q={1 \over 2} \sum_{q, \bar q}
\int_0^1 dx [x q(x) +E_q(x)], 
\eeq
implying, by making use of the momentum and total angular momentum 
conservation, that
\begin{equation}\label{LG}
J_G={1 \over 2} \int_0^1 dx 
[x G(x)-
\sum_{q, \bar q}
E_q(x)],
\eeq
with small $E$ term, qualitatively supports this point of view.

\section{Transverse spin sum rule}

One may wonder, what is a counterpart of a spin sum rule
for the transverse spin case, when twist 3 operators are
involved.
Projecting the Pauli-Lubanski vector on the transverse
direction, the conservation of the total angular momentum leads to
the sum rule
\begin{equation}\label{SRT}
{1 \over 2} \sum_{q, \bar q} S^T_q + S^T_G+L^T_q+L^T_G= {1 \over 2},
\eeq
where transverse spin of quarks and gluons are given by the integrals:
\begin{equation}\label{ST}
S^T_q=\int_0^1 g_q^T(x) dx; \ S^T_G=\int_0^1 \Delta G_T(x) dx.
\eeq
Here $g_T=g_1+g_2$ is the natural measure of the quark contribution to
transverse polarization, as given by the matrix element:
\barr\label{g_T (x)}
g_T (x)=\int {d \lambda \over {4\pi}} e^{i\lambda x}<p,S|\bar \psi(0)
/\!\!\! S_T \gamma_5
 \psi(\lambda n)|p,S>,
\earr
while $\Delta G_T$ is the similar quantity for gluons \cite{ST96}.
Only operators containing two quark or gluon fields appear, while
quark-gluon and three-gluon operators are related to them by the
equations of motion.
Here only chiral even 
quark 
operators contribute , which is the
immediate result of the fact that the quark energy momentum tensor is
chiral even, so that transversity does not enter this sum rule.

The important simplification of the transverse sum rule comes from the
Burkhardt-Cottingham sum rule\cite{BC}, which is valid for each quark
flavour and gluons separately and  states that the first moment of
longitudinal and transverse distributions are equal,

\begin{equation}\label{BC}
S^T_q=S_q; ~~~~~ S^T_G=S_G.
\eeq

This just mean that the contributions of quarks and gluon spin
to the proton spin behave like Lorentz invariant quantities, although
generally speaking they are not \cite{Ji2}. Lorentz invariance comes
from rotational invariance, guaranteed by the Burkhardt-Cottingham sum
rule, and boost invariance.

As an important ingredient of the derivation of the Burkhardt-Cottingham
sum rule is  the locality of the operator, one may doubt the
Lorentz invariance of the quark and gluon orbital momenta, although
their sum is Lorentz invariant due to the total angular momentum
conservation.

\section{Transversity sum rule}

As transversity distributions come from a chirally odd operator with
its first moment (tensor charge) being subject to renormalization,
one may look for a conserved quantity by considering
fractional moments \cite{Sof1}:

\begin{equation}\label{Sh}
\int_0^1 x^{\alpha} h_1(x) dx=const
\eeq

A direct calculation shows that at leading order
$\alpha=-0.345$, while at next-to-leading order
$\alpha=-0.49$ for $Q^2 \sim 10 GeV^2$. 

At the moment, it is difficult to find an interpretation of these
numbers. They should be rather considered as phenomenological 
inputs, defining a conserved quantity. In particular, it is this
quantity which seems to be the natural candidate for low energy
calculations.

\section{Positivity constraints}

Important constraints for the nucleon spin structure
may be derived from the positivity of the density matrix.
Recall that non-diagonal elements of a density matrix are
constrained by positivity as well as its diagonal elements. This enables
to derive inequalities originally proven at the level of
the parton model\cite{Soffer}, which were shown to be
preserved by the QCD $Q^2$ evolution, up to next-to-leading
order\cite{WV,BST}. They read

\begin{equation}\label{in}
|h_1(x)| \le q_+(x).
\end{equation}

\noindent
where $q_+(x)$ is the quark distribution with helicity parallel
to that of the nucleon.

For gluons, the similar bound  \cite{ST97} reads

\begin{equation}\label{Gt}
|\Delta G_T(x)| \le \sqrt{{{G(x) G_L(x}) \over 2}}
\end{equation}
where $G_L$ is the distribution of longitudinally polarized gluons in
nucleon \cite{GI}, for which (\ref{Gt}) provides a lower bound.

Positivity leads also to constraints\cite{pst} for the off-forward
gluon 
distribution $g (x_1,x_2)$:

\begin{equation}\label{geos1}
x^{'} g (x_1,x_2) \le \sqrt {x_1 x_2 g(x_1)g(x_2)}\cdot
\lambda [P(x_1),P(x_2)],
\end{equation}
with  $x_1$, the light-cone fraction of the parton emitted by the
proton target, $x'$, the fraction of the parton absorbed by the scattered
proton (both fractions with respect to the initial proton momentum),
and $x_2 = x'/(1-x+x')$the light-cone fraction
of the absorbed parton  with respect to its parent's momentum, and

\begin{equation}\label{lamb}
\lambda [P(x_1),P(x_2)]=
\frac{\sqrt{(1+P(x_1))
(1+P(x_2))}+\sqrt{(1-P(x_1))(1-P(x_2))}}{2}\nonumber
\end{equation}
where one introduces the gluon polarization, defined as
$P(x)$=$ \Delta G (x)/G(x)$.
This inequality offers, in principle, a possibility of extracting
information on the gluon spin-dependent distribution $ \Delta G$
from unpolarized diffractive processes.

To conclude, let us stress again that sum rules and positivity might 
be used in order to provide constraints on the spin
structure of nucleon, including the most difficult 
transverse polarization case.

\section*{Acknowledgments}
We are thankful to C. Bourrely for assistance.
O.T. is indebted to the Organizers for the opportunity 
to participate in the Conference and warm hospitality 
at Les Arcs and to L. Mankiewicz for valuable
comments.

\end{document}